# Segregation patterns in rotating cylinders determined by the size difference, density ratio, and cylinder diameter

Kurumi Kondo, Hiroyuki Ebata, Shio Inagaki[*]

Department of Physics, Kyushu University, Fukuoka, 819-0395, Japan

[*]Corresponding author
Shio Inagaki: inagaki@phys.kyushu-u.ac.jp

**ABSTRACT**

Granular materials often segregate under mechanical agitation, which differs from the expectation of mixing. It is well known that a bidisperse mixture of granular materials in a partially filled rotating cylinder exhibits alternating bands depending on the combination of the two species. The dynamic angle of repose, which is the angle that a steady avalanche makes with the horizontal, has been considered the dominant parameter that determines the segregated state. However, the previously known angle of repose condition was not always satisfied in different experimental cases. To clarify the experimental conditions, we conducted an exhaustive parameter search with three dimensionless parameters: the particle size difference normalized by the average particle size, the specific density ratio, and the ratio of the cylinder diameter to the average particle size. Additional experiments were conducted to explore the effect of the rotational speed of the cylinder. This systematic approach enabled us to predict the segregated state. Moreover, we discovered that the band width can be effectively scaled by combining these three parameters.

**Introduction**

A mixture of granular particles tends to segregate spontaneously based on size or shape when put in motion by mechanical agitation, such as flow, vibration, or rotation. The dissipative interaction between the particles plays a critical role in preventing mixing, as is generally the case for liquids and gases. In the industrial sector, 75% of raw materials are made of granular material[1]. In terms of environmental issues, demixing technologies are required to remove toxic substances and extract rare metals from urban mines. Thus, efficient mixing and separation technologies are urgently needed. Moreover, the underlying principles must be understood to meet this need.

An intriguing segregation phenomenon arises when a binary mixture of grains in a partially filled horizontal cylinder is agitated by rotation[2,3]. Within a few revolutions, small particles become concentrated in a central core along the axis, and large particles surround this core, which is known as radial segregation. The particles then segregate further into alternating bands along the axis containing either small or large particles. This axial segregation is counterintuitive because inhomogeneous particle distributions are enhanced by rotation despite the uniform mechanical agitation in the axial direction. Since axial segregation was first reported by Oyama in 1939[4], it has been extensively studied through experimental[5-10], theoretical[11-14], and numerical[15-17] methods. Das Gupta *et al.* experimentally showed that axial segregation occurred when $\theta_l < \theta_s$, where $\theta_l$ and $\theta_s$ are the dynamic angles of repose of large and small particles, respectively[18]. The dynamic angle of repose, which is the angle that a steady avalanche makes with the horizontal, has long been widely believed to be the dominant parameter predicting whether axial segregation occurs[18-22].

However, some recent studies have questioned the validity of the condition of the dynamic angle of repose, $\theta_l < \theta_s$, required for axial segregation to occur[23,24]. When radial segregation transforms into axial segregation, if the radial core is completely covered with an outer layer, it is very unlikely that the axial segregation process is driven by a surface phenomenon such as the difference in the dynamic angles of repose. For example, Pohlman *et al.* experimentally showed a counterexample to this empirical condition[23]. Although it cannot be a coincidence that this condition has held in many previous studies, no study has quantitatively determined when this condition holds. Some previous studies have suggested that axial band formation might be caused by the bulging of the radial core of small particles[25-29]. In the process of axial segregation, axial band formation should be considered in two stages: the transition from mixed to radial segregation and the transition from radial to axial segregation. We aim to clarify the necessary conditions for each stage and elucidate the segregation mechanism of granular materials.

## Results

In previous studies, various parameters have been investigated as potential important variables, but in most cases, only one or two parameters were changed, while the others were fixed during experiments[29-34]. In our study, we conducted experiments by exhaustively changing four parameters that are often considered important: the particle size and specific density and the diameter and rotational speed of the cylinder. First, we show segregated patterns with two combinations of particles, both of which are inconsistent with the axial segregation condition of $\theta_l < \theta_s$. The first combination includes glass beads (diameter of the large particle, $d_l$ =2 mm) and zircon beads (diameter of the small particle, $d_s$ = 1.1 mm). The detailed method for measuring the dynamic angle of repose is presented in the Methods section. Here, the rotational speed $\Omega$ and the diameter of the cylinder $D$ are fixed at $\Omega$ = 14 rpm and $D$ = 74 mm, respectively. As shown in Fig. 1 (a), a mixture of glass beads and zircon beads does not show axial segregation, even though the condition of $\theta_l < \theta_s$ is satisfied at all rotation rates in the measurement range. Radial segregation can be observed from the side view (not shown). Although the cylinder was rotated for more than a week, axial segregation did not occur. The second combination includes alumina beads ($d_l$ =2 mm) and glass beads ($d_s$ = 0.6 mm). This mixture shows clear axial segregation (Fig. 1 (b)), even though there is no significant difference in the dynamic angles of repose. This result indicates that the condition $\theta_l < \theta_s$ is not necessarily satisfied when axial segregation occurs.

Figure 1 (c) highlights the contradiction in the dynamic angles of repose in the phase diagram of the segregated state. In this figure, we plot the specific density ratio, $\rho^* = \rho_l/\rho_s$, against the difference in the dynamic angles of repose, $\theta_s - \theta_l$. Note that $\rho_l$ and $\rho_s$ represent the specific densities of the large and small particles, respectively. Experiments were conducted with different combinations of four types of particles: glass, alumina, zircon, and zirconia beads. The rotational speed and the diameter of the cylinder were fixed at $\Omega$ = 14 rpm and $D$ = 74 mm, respectively, as in the earlier experiments. According to the heuristic condition for axial segregation, axial segregation should occur in the upper half of the diagram. However, axial segregation appears primarily in the domain $\rho^* > 1$ irrespective of $\theta_s - \theta_l$. Notably, the heuristic condition tends to hold along the dotted line $\rho^* = 1$, as observed in many previous studies. Although the underlying mechanism is not yet fully understood, $\theta_l < \theta_s$ has long been believed to be the condition for axial segregation because previous studies confirming this condition have mainly employed particles with identical specific densities[18,19]. Figure 1 (c) also suggests that the specific density ratio is an important parameter that influences the segregation pattern. However, in the domain $\rho^* > 1$, all three states, namely, radial, mixed, and axial, appear. Thus, to accurately classify the segregation states, we need to consider additional parameters.

Next, we consider the effect of different particle sizes on the segregated states. The diameter of the denser particles, alumina beads, is fixed at 2 mm, while the diameter of the lighter particles, glass beads, is varied from 0.4 to 4 mm. The diameters of the alumina and glass beads are denoted as $d_A$ and $d_G$, respectively. Figure 2 depicts the top views of the cylinder after rotating for 10 minutes. In the case of $d_G > d_A$, the side view shows that radial segregation occurred, with the glass beads covering the core composed of the alumina beads. This segregated radial structure was maintained for more than three days of rotation. This corresponds to the region $\rho^* < 1$ in Fig. 1 (c). In the case of $d_G = d_A$, the two species with nearly identical radii remained in the mixed state for more than 24 hours of rotation. In the case of $d_G < d_A$, after rotating for a few minutes, radial segregation occurred, with the alumina beads covering the core composed of the glass beads. Then, the exterior layer of alumina beads split into alternating bands along the axis of the cylinder. As the size difference increased, the band width became wider. The results indicate that the particle size difference should be considered as a controlling parameter that influences the segregated state.

Fig. 3 (a) demonstrates the effect of the specific density ratio $\rho^*$ and particle size difference on the segregation state. The particle size difference is nondimensionalized as $\Delta^* = (d_l - d_s)/d_{av}$, where the average diameter of the large and small particles is defined as $d_{av} = (d_l + d_s)/2$. In the experiments for this phase diagram, the particle diameter and density, and the cylinder diameter, $D$, were varied while maintaining the rotational speed of $\Omega = 14$ rpm. The axial segregation state and other states (mixed and radial segregation) are mostly separated except near the boundaries. When the particle size difference is zero, segregation does not occur, even for glass and zirconia beads ($\rho^* = 2.44$), the combination with the highest specific density ratio. When $0.3 < \Delta^* \lesssim 1.3$, radial segregation always occurs within a few minutes after the start of rotation. Large particles tend to be present in the external layer while flowing down the surface, which causes radial segregation. When the external layer is composed of less dense larger particles ($\rho^* < 1$), the segregated state remains radial regardless of how long the cylinder is rotated. If $\rho^* \geqq 1$, radial segregation transforms into axial segregation. When $\Delta^* \gtrsim 1.3$ and $\rho^* > 1$, the large particles with high density sink into the axial core of the small particles with low density, as seen in the reverse radial segregation region[35], leading to a mixed or radial state in which the core is composed of large particles with high density. Figure 3 (a) demonstrates that axial bands are formed only when the external layer is composed of larger particles with high density and the size difference is moderate.

A critical issue in the phase diagram in Fig. 3 (a) is that the segregated states cannot be completely separated along the boundaries between the axial segregation domain and the other domains ($\rho^* \sim 1$ or $\Delta^* \sim 1.3$). When $\Delta^* < 1.3$, the regions with axial and radial patterns can be mostly separated based on the density ratio. However, along the line of $\rho^* = 1$, both patterns appear with the same particle size difference. Alexander *et al.*[29] reported that axial segregation disappeared when the ratio of the cylinder diameter to the average particle size, $D^* = D/d_{av}$, was less than a certain threshold. This suggests that in our phase diagram, an axis related to $D^*$ should be added to accurately classify the segregated states.

Figure 3 (b) depicts a three-dimensional phase diagram with $\Delta^*$, $1/D^*$, and $\rho^*$. For visibility, the data points with small particle size differences ($\Delta^* < 0.5$) are not plotted. The domains of axial segregation and the other states can be completely separated by two planes. The procedure to determine the planes is provided in the Methods section. To ensure that the boundaries between the different phases are clearly visible, we plot a two-dimensional phase diagram, as shown in Fig. 3 (c). First, the three-dimensional phase diagram is linearly transformed so that the two planes are orthogonal. Next, the rescaled two-dimensional phase diagram is drawn using the new axes $\widehat{\Delta}^*$ and $\widehat{\rho}^*$. These new axes are defined as

$$\widehat{\Delta}^* = \Delta^* + \beta_1 \rho^* + \frac{\alpha_1}{D^*}, \tag{1}$$

$$\widehat{\rho}^* = \rho^* + \beta_2 \Delta^* - \frac{\alpha_2}{D^*} \tag{2}$$

where the fitting parameters $\alpha_i$ and $\beta_i$ were calculated based on the equation of the separating plane in Fig. 3(b). The values of $\alpha_i$ and $\beta_i$ are listed in Table 1. Then, we found that $\alpha_1 \sim \alpha_2$ and $\beta_1 \sim \beta_2$. Note that the separating lines in Fig. 3 (c), $\widehat{\rho}^* = \rho_0 \sim 1$ and $\widehat{\Delta}^* = \gamma_2$ correspond to the separating planes in the original 3D phase diagram (Fig. 3 (b)). The type of segregation depends mainly on the density ratio and the size difference, which are represented by the first terms on the right-hand sides of Eqs. (1) and (2). The remaining terms, which are on the order of 0.1 or even less, can be regarded as correction terms. Based on the dependence of the phase diagram on the rotational speed of the cylinder, qualitatively similar phase diagrams were obtained for different rotational speeds, as demonstrated in Fig. S1 in the Supplemental Materials. The only apparent differences in the phase diagrams were the values of $\alpha_i$ and $\beta_i$, as shown in Table 1. Thus, we found that the final segregated state is determined primarily by $\Delta^*$ and $\rho^*$, while the correction terms and $\Omega$ play crucial roles only near the boundaries in the phase diagram.

The final segregated state is governed by three key parameters: $D^*$, $\Delta^*$, and $\rho^*$. These parameters dictate the extent to which the system transitions from its initially mixed state to radial segregation and further evolves into axial segregation. It is anticipated that these three parameters also exert an influence on the dynamics of band formation. Here, we explore the relationship between the initial band width and these three parameters. As shown in Fig. 4, the dimensionless initial wavelength of the bands, $\lambda^* = \lambda/d_s$, was found to be linearly proportional to the combination of the dimensionless values $D^*\Delta^*/\rho^*$ (see the definition of $\lambda$ in the Methods section). Among the various combinations that were investigated, the most distinct linear trend was observed when the wavelength was scaled with the diameter of the small particles, $d_s$. Figure 4(a) illustrates the linear dependence of the dimensionless initial wavelength on not only the dimensionless size difference, $\Delta^*$ but also the ratio of the cylinder diameter to the average particle size, $D^*$, and the inverse of the specific density ratio, $1/\rho^*$. This is consistent with previous studies showing that the band width is proportional to the diameter of a cylinder with fixed $\Delta^*$ and $\rho^{*}$[31,35]. We emphasize that even when we varied the rotational speed from 8 to 34 rpm, all of the data points appeared to fall on the same line. The changes in the rotational speed did not lead to any systematic trends with increasing or decreasing $\lambda^*$ (Fig. 4(b)). Given that we investigated the very early stage of band formation, our findings indicate that the dimensionless initial wavelength is not significantly affected by the rotational speed in our experiments. However, the dimensionless parameters $D^*$, $\Delta^*$, and $\rho^*$ play dominant roles in determining the initial wavelength. The unscaled wavelength, $\lambda$, was also plotted against the rotational speed, as shown in Fig. S2(a) in the Supplemental Materials. As the rotational speed increased, the wavelengths of certain parameter combinations increased, while others decreased. The variations in the initial wavelength due to $D^*\Delta^*/\rho^*$ are more pronounced when the wavelength is scaled by $d_s$ compared to the variations when the initial wavelength depends on $\Omega$.

**Discussion**

As reported in previous studies, radial segregation is considered an essential precursor of axial segregation[9,44]. Some previous studies have suggested that the formation of axial bands could be attributed to the bulging of the radial core consisting of small particles[25-29]. The conditions for axial segregation to occur should be considered in two stages: the transition from mixed to radial segregation and the transition from radial to axial segregation. The phase diagram projected onto a two-dimensional space in Fig. 3(c) revealed that the former condition corresponds to $\gamma_1 < \widehat{\Delta}^* < \gamma_2$, while the latter condition corresponds to $\hat{\rho}^* > \rho_0$. We emphasize that these inequalities are described not merely by a single parameter such as the size difference or the density ratio but by a linear combination of three parameters, $D^*$, $\Delta^*$, and $\rho^*$, as defined in Eqs. (1) and (2). In the transition from mixed to radial segregation, the size difference should exceed a certain threshold. However, if the size difference is too large, the larger particles form a core, which is sometimes called reverse segregation[36]. Within the investigated parameter range, reverse radial segregation never transformed into axial segregation. Therefore, to initiate radial segregation as a precursor to axial segregation, the size difference, including the correction terms, must have both upper and lower limits. The transition from radial to axial segregation is primarily determined by the specific density ratio. Radial segregation with a core consisting of small particles tends to transform into axial segregation, particularly when the large particles have a higher specific density than the small particles. Since there are correction terms in Eq. (2), the density ratio itself does not have to be greater than 1. A more precise lower limit for the specific density ratio is discussed later. At the onset of axial segregation, the external layer consisting of large particles with higher specific density on top of the core of small particles with lower specific density destabilizes and splits into axial bands, which is similar to Rayleigh-Taylor instability. D'Ortona and Thomas recently demonstrated Rayleigh-Taylor instability between two dry granular materials of different densities[20]. Alternating bands emerge as the superimposed layers continue to flow. It would be interesting to determine how the wavelength changes with $\Delta^*$, $\rho^*$, and the system size in their experiments as a comparison with our results.

We next closely examine the conditions for axial segregation illustrated in Fig. 3(c). The dotted lines delineating the boundaries between the axial segregation state and the other states suggests two conditions: $\gamma_1 < \widehat{\Delta}^* < \gamma_2$ and $\hat{\rho}^* > \rho_0$. By transforming these inequalities, we derive the following simultaneous inequality:

$$\frac{1}{\beta_2}\{\alpha_2 + (\rho_0 - \rho^*)D^*\} < D^* \Delta^* < (\gamma_2 - \beta_1 \rho^*)D^* - \alpha_1. \tag{3}$$

Bielenberg et al.[32] experimentally found that axial segregation occurred when $D^* \Delta^*$ was larger than a certain threshold value. Equation (3) suggests that there also exists an upper threshold of $D^* \Delta^*$ that leads to axial segregation and that the upper and lower thresholds should depend on both $\rho^*$ and $D^*$. For Eq. (3) to be valid, the left-most side of the inequality must be smaller than the right-most side. This requirement places a necessary condition for axial segregation on $D^*$:

$$D^* > \frac{\alpha_1 \beta_2 + \alpha_2}{\gamma_2 \beta_2 - \rho_0 + (1 - \beta_1 \beta_2)\rho^*}. \tag{4}$$

This lower limit of $D^*$ for axial segregation to occur is consistent with the lower limits obtained in previous studies based on combinations of different-sized glass beads ($\rho^* = 1$) under similar rotational speeds[29,31]. Equation (4) is an extension of the lower limit of $D^*$, where axial segregation occurs, for an arbitrary combination of specific densities. This equation also provides the lower limit of the specific density ratio required for axial segregation. The denominator on the right-hand side of Eq. (4) must be positive. This leads to the condition that $\rho^* \geq 0.5$ for a rotational speed of 14 rpm. Thus, axial segregation occurs when the large particles have a specific density at least 0.5 times that of the small particles at $\Omega = 14$ rpm. As the rotational speed varies, $\alpha_i$ and $\beta_i$ change, which affects the threshold for the density ratio required to initiate axial segregation.

The phase diagram shown in Fig. 3(c) offers new insights into the results of previous studies. Hill and Kakalios demonstrated reversible axial segregation by exploiting the different dependencies of the dynamic angles of repose on rotational speed for glass beads of different sizes[19]. Since $\rho^* = 1$ in their study, the parameters were located near the boundary between axial and radial segregation in their phase diagram. Reversible axial segregation can be interpreted as occurring when the boundary between axial and radial segregation shifts reversibly with the rotational speed. We obtained qualitatively similar phase diagrams for different rotational speeds, as shown in Fig. S1 in the Supplemental Materials, with the only apparent differences being the values of $\alpha_i$ and $\beta_i$. Thus, the rotational speed affects the segregated state only near the phase boundaries. A similar argument can be made for changing $D^{*}$[29]. As mentioned earlier, previous experimental results have shown that reducing $D^*$ with a fixed value of $\rho^* = 1$ led to the transition from the axial to the radial segregation state. This can be attributed to the fact that the magnitude of the third term on the right-hand side of Eq. (2) increased, causing $\hat{\rho}^*$ to cross the boundary from the axial to the radial segregation region in the phase diagram. As the third term is typically on the order of 0.1 or less, their parameters must have been near the boundary. Moreover, near the transition from axial to radial segregation upon reducing $D^*$, they observed reversible axial segregation in response to rotational speed changes. This provides additional evidence to support our interpretation of the phase diagram. Thus, the transition between the axial and radial segregation states observed in previous studies can be understood in terms of a phase diagram including $D^*$, $\Delta^*$, and $\rho^*$. Furthermore, axial segregation should not disappear if $\rho^*$ is sufficiently large far from the boundary as $D^*$ and $\Omega$ change.

Although several previous studies have examined the relationship between the band width and rotational speed[33,35,37-40], there is no general agreement as to whether the band width increases or decreases with increasing rotational speed. Furthermore, to our knowledge, no experiments have been conducted to explore the relationship between the band width and specific density ratio. Although this linear relation shown in Fig. 4(a) lacks a clear underlying mechanism, it enables us to predict the band width for any combination of spherical particles using $D^*$, $\Delta^*$, and $\rho^*$ as long as the parameters remain in the axial segregation region in the phase diagram. The inverse proportionality between the wavelength and the specific density ratio is similar to Rayleigh-Taylor instability, which may elucidate the fundamental mechanism of band formation.

According to linear stability analyses of pattern formation, when staring with an initially random (mixed) state, the most unstable wavelength dominantly grows when the perturbation is sufficiently small. Thus, Fig. 4 suggests that the most unstable wavelength depends on $D^*$, $\Delta^*$, and $\rho^*$.

The prominent linear trend that was observed when scaling $\lambda$ by $d_s$ may be elucidated by considering the process of axial segregation. Some previous studies[25-29] have suggested that following radial segregation, a radial core may protrude periodically along the axis, driven by shear from a surface avalanche. Subsequently, the external layer splits to form axial bands, with the initial wavelength corresponding to the length of one bulge. As the core is mainly composed of small particles, we suggest that the length of the bulge could be determined as an integer multiple of the diameter of a small particle, rather than the average diameter of the large and small particles. To validate our hypothesis, further experiments are needed, which would elucidate the underlying mechanisms of axial segregation.

Several previous studies have assessed the influence of end walls on axial band formation[24, 41,42]. Bands rich in large particles tend to initially form at the end walls of the cylinder. Pohlman *et al.* measured the surface velocity in a monodisperse system using particle tracking velocimetry and observed axial flow near the end walls[43]. If the cylinder length is not sufficiently large relative to its diameter, the walls may have a significant effect on the results. However, Fiedor *et al.* showed experimentally that the number of bands per unit length is independent of the cylinder length when the cylinder length is varied while keeping the diameter unchanged[44]. Therefore, we suggest that band formation is primarily caused by global factors, such as the bulging of the core, rather than localized effects, such as the presence of end walls.

In the present study, we focused on investigating several key parameters that may determine the axial/radial/mixed states of a binary granular mixture in a rotating cylinder. To comprehensively explore this phenomenon, we systematically varied the following parameters: the diameter and specific density of the particles and the diameter and rotational speed of the cylinder. It is widely known that axial segregation occurs when the dynamic angle of repose of the large particles is less than that of the small particles. A thorough parameter search reveals that few particle combinations satisfy this condition. The size difference normalized by the average diameter, the specific density ratio, and the cylinder diameter normalized by the average particle diameter are proposed to determine the segregated state. An intriguing finding of our study is that the three-dimensional phase diagram effectively categorizes the segregated states based on only the material properties of the particles and the system size at a fixed rotational speed. This discovery underscores the profound influence of these parameters on the emergent patterns, providing valuable insights into the underlying mechanisms governing segregation phenomena in rotating granular systems. Varying the rotational speed distorts the phase diagram, revealing the sensitivity of the segregation behavior to the rotational speed, especially near the boundaries. The parameters that are related to the dynamics, such as the dynamic angle of repose and fluidity, influence the parameters $\alpha_i$, $\beta_i$, and $\gamma_i$, which determine the boundary. However, many parameters associated with segregation phenomena in rotating cylinders were not considered in the present study, such as the fill level[39,45] and the fraction of large to small particles[40]. Further experiments are needed to explore the influence of these parameters on the three-dimensional phase diagram of the segregated state discovered in this study.

## Methods
**Experimental setup**

The particles were glass, alumina, zircon, or zirconia beads, whose specific densities are 2.5, 3.8, 4.4, and 6.1 g/cm$^3$, respectively. The diameters of the particles were varied from 0.2 to 4 mm. Acrylic transparent cylinders with three different inner diameters (*D*) were used: 36, 54, and 74 mm. The lengths of the cylinders (*L*) were fixed at 320 mm. Half the volume of the cylinder was filled with an equal-volume mixture of two types of particles. The cylinder was placed horizontally on two shafts parallel to its axis and rotated at a constant rotational speed *Ω* by driving one of the shafts with a brushless DC electric motor (Oriental Motors). The rotational speed was varied from 8 to 34 rpm. We performed experiments with 326 parameter combinations.

**Measurement method of the dynamic angle of repose**

The dynamic angle of repose of each species was measured using a half-filled rotating cylinder with a single species (upper panel of Figs. 1 (a, b)). For the examined rotational speeds, the surface of the granular bed was nearly flat, as shown in Fig. S3 in the Supplemental Materials. Then, we imaged the avalanche from the side view and measured the slope of the surface. All the represented points were obtained by averaging over five independent measurements. The error bars correspond to the standard error of the mean.

**Parameter estimation of the separating plane by using a support vector machine**

The two planes in the $\Delta^* - 1/D^* - \rho^*$ phase diagram were determined with the following procedure. Except for the mixed state at small $\Delta^*$, the axial segregation state and other states were roughly distinguished by the conditions $\Delta^* \sim 1.3$ and $\rho^* \sim 1$ (Fig. 3(b)). To determine the separating plane near $\rho^* \sim 1$, we selected the data points satisfying $\rho^* \leq 1.2$ and $\Delta^* \leq 1.4$. To determine the separating plane near $\Delta^* \sim 1.3$, we used the data satisfying $\rho^* > 1$ and $\Delta^* > 1.2$. Then, we calculated the separating planes $a_i \Delta^* + b_i \rho^* + c_i/D^* + 1 = 0$ (i=1,2) using a support vector machine (SVM) with a linear kernel function and infinitely large box constraint (MATLAB software). To calculate the standard deviation of the coefficients $a_i$, $b_i$, and $c_i$, random values were introduced to each coefficient obtained with the SVM, and the resulting data points that were consistent with the established boundaries were collected. Based on the collected data, the averages and standard deviations of the coefficients were computed, as listed in Table 1. Equation (1) is defined as $\widehat{\Delta}^* = \Delta^* + \beta_1 \rho^* + \alpha_1/D^*$, with $\alpha_1 = c_1/a_1$, and $\beta_1 = b_1/a_1$. Equation Eq. (2) is defined as $\hat{\rho}^* = \rho^* + \beta_2 \Delta^* - \alpha_2/D^*$ with $\alpha_2 = -c_2/b_2$, and $\beta_2 = a_2/b_2$. The threshold values were calculated as $\gamma_2 = -1/a_1$ and $\rho_0 = -1/b_2$. While the precise estimation of $\gamma_1$ was limited due to insufficient data, it was consistently calculated as approximately 0.6 for all tested rotation speeds. In Fig. 4 (c) and Fig. S1 in the Supplemental Material, we assigned $\gamma_1$ a fixed value of 0.6.

**Definition of the wavelength of the axial bands**
We counted the initial number of bands that appeared within the first 5 minutes of rotation. Trials in which bands took longer than 5 minutes to appear were excluded because merging may have occurred in some regions in the experiments under these conditions. Since the ends of the cylinder are usually occupied by the bands consisting of large particles, the initial wavelength ($\lambda$) is defined as $L / (N_0 + 1/2)$, where $L$ is the length of the cylinder and $N_0$ is the initial number of bands rich in small particles.

**Acknowledgments** We thank Michio Otsuki for his careful reading and helpful comments. This work was supported by JSPS KAKENHI Grant Numbers 18K03463, 19K14614, and 22K03468.

**Author contributions** H. E. and S. I. conducted all derivations, computations, and analyses. K. K. and S. I. conducted segregation experiments. S. I. proposed the project and directed the research, and proposed the analyses. H. E. and S. I. discussed the results and wrote the paper.

**Competing financial interests** The authors declare no competing interests.

**Data and materials availability** All data needed to evaluate the conclusions in the paper are present in the paper and/or the Supplementary Materials. Additional data related to this paper may be requested from the authors.

**Author information** Correspondence and requests for materials should be addressed to S.I. (inagaki@phys.kyushu-u.ac.jp).

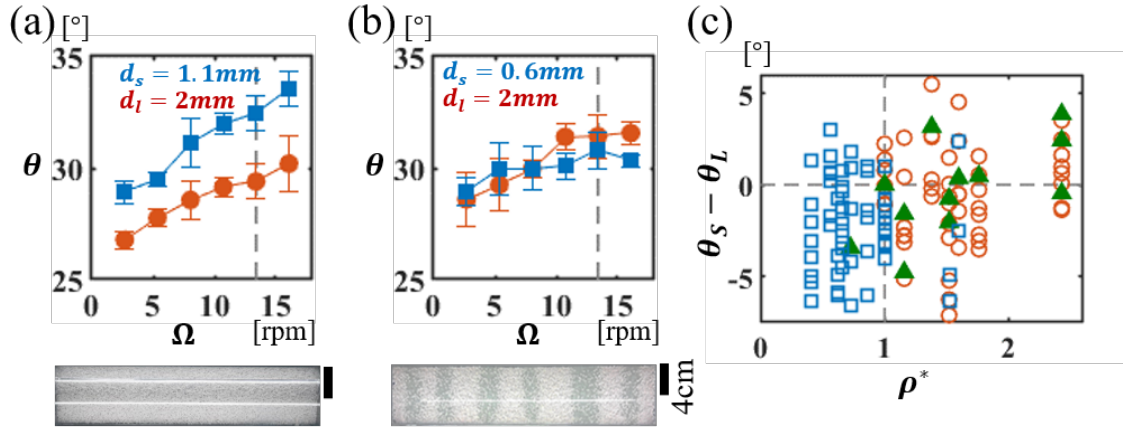

**Figure 1.** (a, b) Upper panels: Dynamic angle of repose of a single species as a function of the rotational speed ($\Omega$). The dashed line shows $\Omega$ at 14 rpm. Lower panels: Top views of the cylinder after 10 minutes of rotation ($\Omega$ = 14 rpm). (a) Glass beads ($d_l$ = 2 mm, red circles) and zircon beads ($d_s$ = 1.1 mm, blue squares). (b) Alumina beads ($d_l$ = 2 mm, red circles) and glass beads ($d_s$ = 0.6 mm, blue squares). (c) Difference in the dynamic angles of repose, $\theta_s - \theta_l$, vs. the specific density ratio, $\rho^* = \rho_l/\rho_s$. Three segregated states: radial (blue squares), mixed (green filled triangle), and axial (red circles) states with various combinations of particles with specific densities and diameters. Combinations of spherical particles are chosen from 4 types of materials (glass, alumina, zircon, and zirconia), and the diameters range from 0.2 to 4.0 mm. ($\Omega$ = 14 rpm and $D$ = 74 mm.)

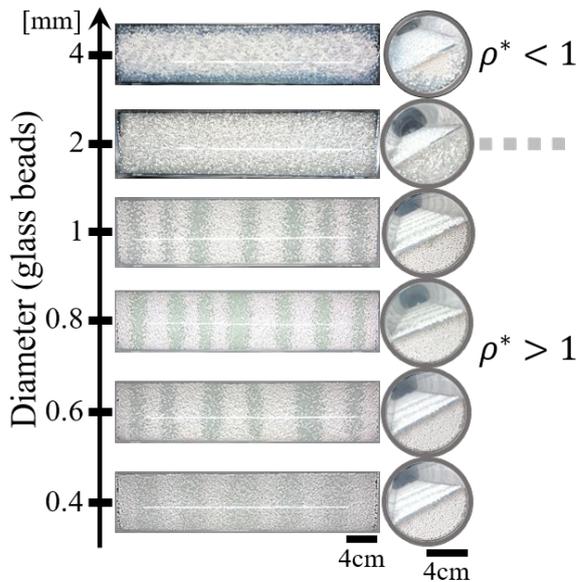

**Figure 2.** Segregated state after 10 minutes of rotation ($\Omega$ = 14 rpm). The white alumina beads have a diameter of 2 mm. The diameters of the glass beads (gray) are varied from 0.4 to 4 mm. Left panels: Top views of the cylinder. Right panels: Side views of the cylinder.

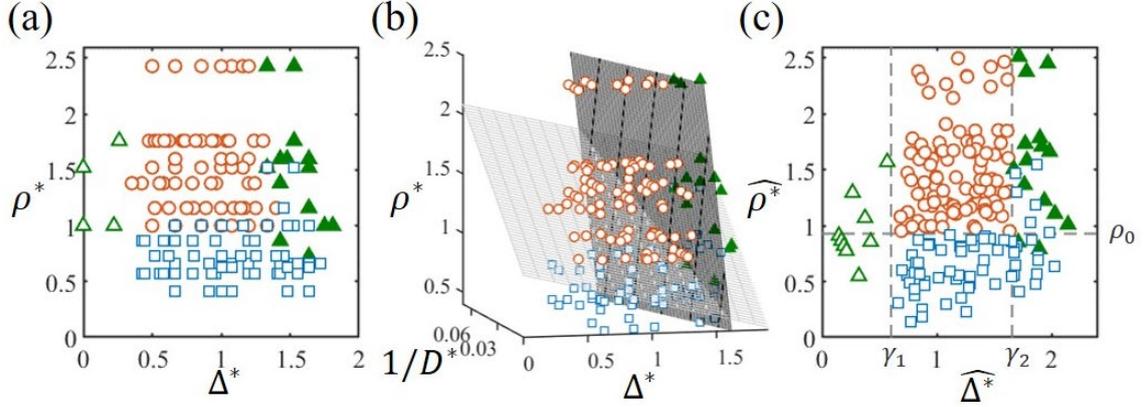

**Figure 3.** Phase diagrams of the segregated states: radial (blue squares), mixed ($\Delta^* > 0.5$, green filled triangle), mixed ($\Delta^* < 0.5$, green open triangle), and axial (red circles). Mixed states with $\Delta^* < 0.5$ and $\Delta^* > 0.5$ are distinguished and marked differently, reflecting their distinct origins of mixing. (a) Phase diagram using $\Delta^* = (d_l - d_s)/d_{av}$ and $\rho^* = \rho_l/\rho_s$. (b) 3D phase diagram using $\Delta^*$, $\rho^*$, and $1/D^*$. The mesh planes indicate the separating planes. For visibility, the mixed state at $\Delta^* < 0.5$ is not plotted. (c) Phase diagram projected onto 2D space, with $\widehat{\Delta}^*$ and $\widehat{\rho}^*$ as axes, according to Eqs. (1) and (2). The dotted lines correspond to the boundaries of the segregated state, with $\gamma_1$ and $\gamma_2$ indicating the positions of the vertical dotted lines and $\rho_0$ indicating the position of the horizontal dotted line. The combinations of the particle species are chosen from glass, alumina, zircon, and zirconia beads with sizes ranging from 0.2 mm to 4 mm. Three different inner cylinder diameters ($D$) are used: 36 mm, 54 mm, and 74 mm. Here, 184 parameter combinations are considered.

**Figure 4.** (a) The normalized initial wavelength ($\lambda^*$) vs. the dimensionless parameter ($D^* \Delta^*/\rho^*$) for different rotational speeds of 8 rpm (purple circles), 14 rpm (yellow stars), 20 rpm (green triangles), 27 rpm (blue squares), and 34 rpm (red inverted triangles). (b) The normalized initial wavelength ($\lambda^*$) vs. the rotational speed ($\Omega$).

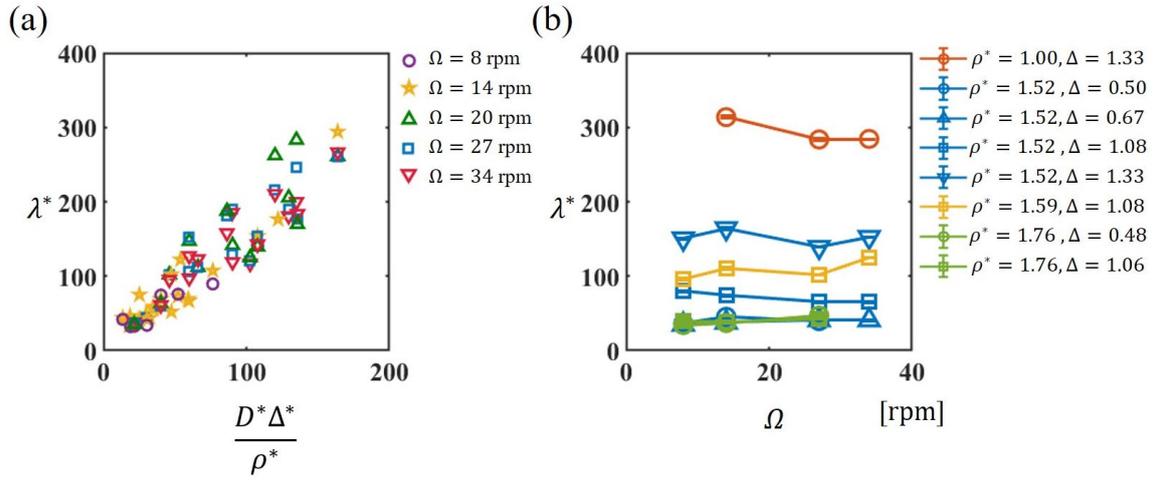

**Table 1.** List of the coefficients used in Eqs. (1) and (2), $\alpha_i$, $\beta_i$, estimated using an SVM model.

| $\Omega$ (rpm) | $\alpha_1$ | $\beta_1$ | $\alpha_2$ | $\beta_2$ |
|---|---|---|---|---|
| 8 | 14.6 ± 2.9 | 0.13 ± 0.02 | 19.3 ± 2.4 | 0.04 ± 0.01 |
| 14 | 8.0 ± 1.2 | 0.10 ± 0.01 | 16.6 ± 2.0 | 0.27 ± 0.03 |
| 27 | 0.0 ± 0.0 | 0.18 ± 0.05 | 13.4 ± 2.4 | 0.19 ± 0.06 |

## A. Dependence of phase diagrams on the rotational speed

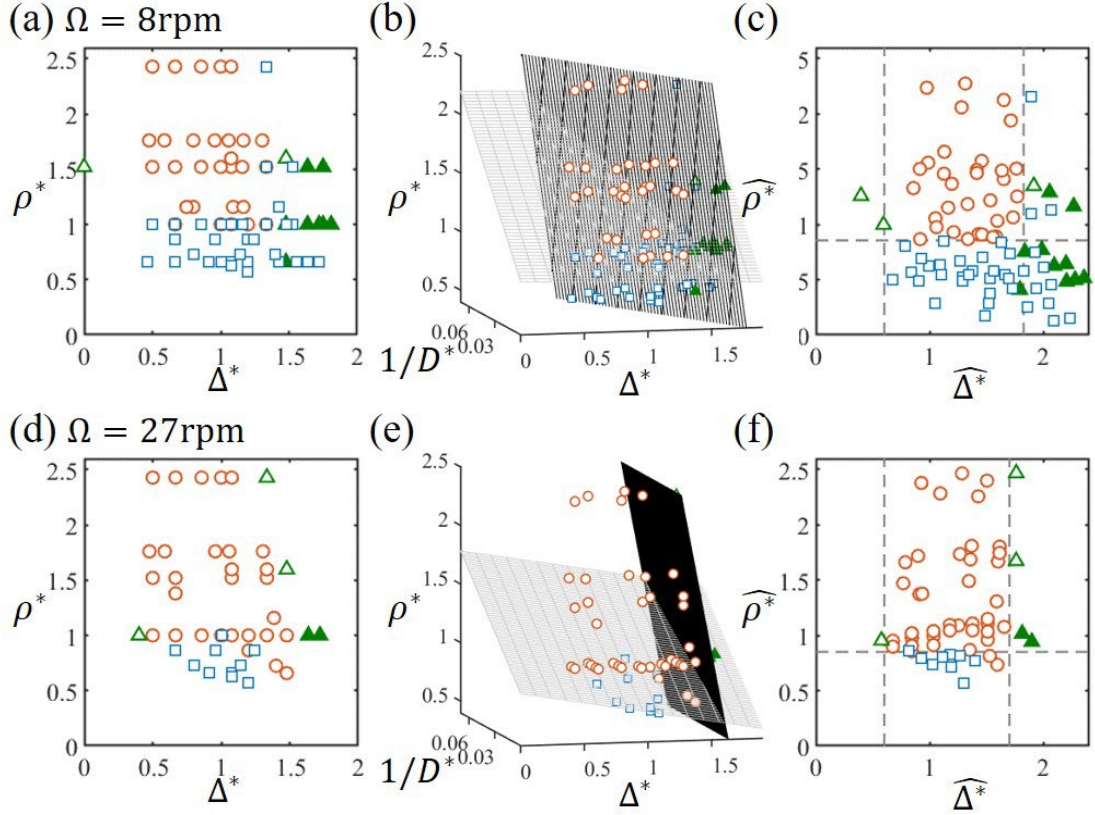

Figure S1. Phase diagrams of the segregated states: radial (blue squares), mixed ($\Delta^* > 0.5$, green filled triangle), mixed ($\Delta^* < 0.5$, green open triangle), and axial (red circles). Mixed states with $\Delta^* < 0.5$ and $\Delta^* > 0.5$ are distinguished and marked differently, reflecting their distinct origins of mixing. (a-c) $\Omega = 8$ rpm (d-f) $\Omega = 27$ rpm (a, d) Phase diagram using $\Delta^* = (d_l - d_s)/d_{av}$ and $\rho^* = \rho_l/\rho_s$. (b, e) 3D phase diagram using $\Delta^*$, $\rho^*$, and $1/D^*$. The mesh planes indicate the separating planes. For visibility, the mixed state at $\Delta^* < 0.5$ is not plotted. (c, f) Phase diagram projected onto 2D space, with $\widehat{\Delta}^*$ and $\widehat{\rho}^*$ as axes, according to Eqs. (1) and (2). The dotted lines correspond to the boundaries of the segregated state. The combinations of the particle species are chosen from glass, alumina, zircon, and zirconia beads, with diameters ranging from 0.2 mm to 4 mm. Three different inner cylinder diameters ($D$) are used: 74 mm, 54 mm, and 36 mm. Here, 86 parameter combinations were considered for $\Omega = 8$ rpm, and 56 parameter combinations were considered for $\Omega = 27$ rpm.

To verify the influence of the rotational speed on the phase diagram of the segregated state, we conducted experiments at different rotational speeds, as shown in Fig. S1. We obtained qualitatively similar phase diagrams. The only apparent differences in the phase diagrams were the values of $\alpha_i$ and $\beta_i$. The parameters to determine the separating planes between each phase are shown in Table 1.

## B. Dependence of the initial wavelength on the rotational speed

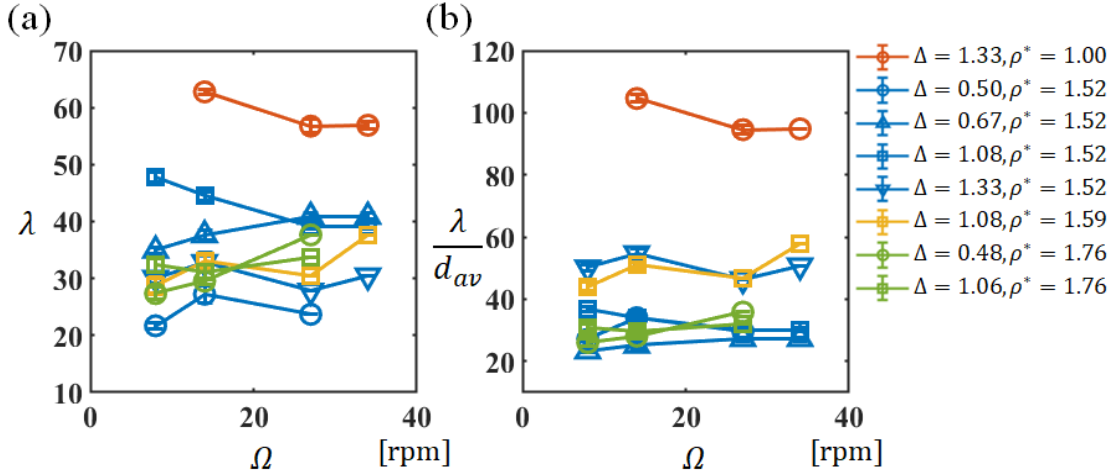

Figure S2. (a) Initial wavelength ($\lambda$) vs. rotational speed ($\Omega$) for various particle combination. (b) Initial wavelength scaled by the average diameter ($\lambda/d_{av}$) vs. the rotational speed ($\Omega$) with various particle combinations. The cylinder diameter is fixed at 74 mm.

The dependence of the initial wavelength on the rotational speed is shown in Fig. S2. The unscaled wavelength, $\lambda$, is plotted against $\Omega$ in Fig. S2(a) for various size differences and density ratios. No significant trends in $\lambda$ with respect to changes in $\Omega$, $\Delta^*$ or $\rho^*$ were observed. Fig. S2(b) presents the initial wavelength scaled by the average diameter of the large and small particles as a function of $\Omega$. The ranges of $\lambda$ and $\lambda/d_{av}$ increasing and decreasing with the rotational speed is wider than that of $\lambda/d_s$, as shown in Fig. 4(b). Even when plotting $\lambda$ and $\lambda/d_{av}$ against $D^* \Delta^*/\rho^*$, a positive correlation is observed, but as shown in Fig. 4(a), plotting $\lambda/d_s$ against $D^* \Delta^*/\rho^*$ yields a pronounced linear relationship with a higher correlation coefficient.

## C. Shape of the surface flow at the end wall of a cylinder

The shapes of the surface avalanche for different rotational speeds are shown in Fig. S3 for glass beads and alumina beads. Smaller grain sizes slightly distort the shape of the surface flow, but the shapes of the surface avalanche can be regarded as flat within the range of our experiments.

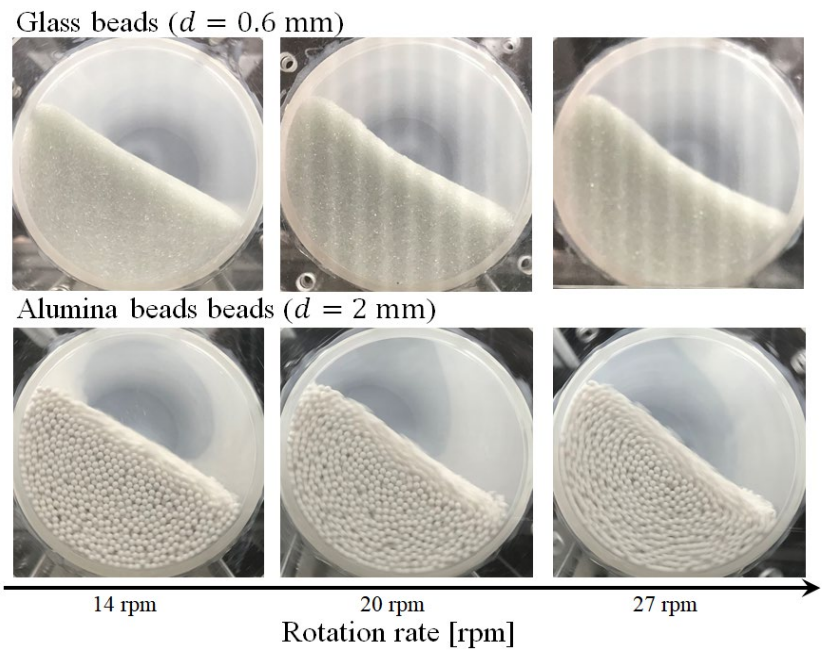

Figure S3. Shape of the surface flow from the side view.